\documentclass[conference,10pt]{IEEEtran}
\IEEEoverridecommandlockouts
\usepackage{booktabs} % For formal tables
\usepackage{url} % Referencing URLs in document
\usepackage{graphicx} % Add graphics to the document
\usepackage{amsmath,amssymb,amsfonts} % Insert equations
\usepackage{xcolor} % Import varied colors into the document
\usepackage{caption} % Option to customize captions for figures, and floats\
\usepackage{textcomp} % Used for baht, bullet (lists), copyright, etc.
\usepackage{dblfloatfix}
\usepackage{fancyhdr} % Used to insert headers into the document by changing page style

\usepackage{soul} % Used for spacing out, underlining, striking out, highlighting, etc/
\usepackage{enumitem} % Complete control of three list environments: enumerate, itemize, and description
\usepackage{lipsum} % Generates garbage text for checking
\usepackage[hang,flushmargin]{footmisc}
\usepackage{etoolbox}
\usepackage{float}
\usepackage{multirow}
\usepackage{hhline}

\newlist{inlinelist}{enumerate*}{1}
\setlist[inlinelist]{label=(\roman*)}

\usepackage{algorithmicx}
\usepackage{algorithm}
\usepackage{algpseudocode}
\usepackage{wrapfig}

\algnewcommand\algorithmicinput{\textbf{Input:}}
\algnewcommand\Input{\item[\algorithmicinput]}

\algnewcommand\algorithmicconst{\textbf{Constraints:}}
\algnewcommand\Const{\item[\algorithmicconst]}

\algnewcommand\algorithmicoutput{\textbf{Output:}}
\algnewcommand\Output{\item[\algorithmicoutput]}

\algnewcommand{\algorithmicgoto}{\textbf{go to}}%
\algnewcommand{\Goto}[1]{\algorithmicgoto~\ref{#1}}%

\algrenewcommand\algorithmicindent{0.5em}

\renewcommand{\thefigure}{\arabic{figure}}

\usepackage{cite}
\usepackage{array}
\newcolumntype{L}[1]{>{\raggedright\let\newline\\\arraybackslash\hspace{0pt}}m{#1}}
\newcolumntype{C}[1]{>{\centering\let\newline\\\arraybackslash\hspace{0pt}}m{#1}}
\newcolumntype{R}[1]{>{\raggedleft\let\newline\\\arraybackslash\hspace{0pt}}m{#1}}
\newcolumntype{M}[1]{>{\centering\arraybackslash}m{#1}}
\newcolumntype{O}[1]{>{\raggedleft\arraybackslash}m{#1}}
\usepackage{makecell}
\usepackage{diagbox}

\usepackage[hidelinks, bookmarks=false]{hyperref}

\newcommand{\printfnsymbol}[1]{%
  \textsuperscript{\@fnsymbol{#1}}%
}

\def\BibTeX{{\rm B\kern-.05em{\sc i\kern-.025em b}\kern-.08em
    T\kern-.1667em\lower.7ex\hbox{E}\kern-.125emX}}

\begin{document}

\title{\huge ApproxFPGAs: Embracing ASIC-Based Approximate Arithmetic Components for FPGA-Based Systems}

\author{
    \IEEEauthorblockN{
    Bharath Srinivas Prabakaran\IEEEauthorrefmark{1}\textsuperscript{,}\IEEEauthorrefmark{3}\thanks{\IEEEauthorrefmark{3}~These two authors have contributed to this work equally.}, Vojtech Mrazek\IEEEauthorrefmark{2}\textsuperscript{,}\IEEEauthorrefmark{3}, Zdenek Vasicek\IEEEauthorrefmark{2}, Lukas Sekanina\IEEEauthorrefmark{2}, Muhammad Shafique\IEEEauthorrefmark{1}}
    \IEEEauthorblockA{\IEEEauthorrefmark{1}Institute of Computer Engineering, Technische Universit{\"a}t Wien (TU Wien), Austria
    \\\{bharath.prabakaran, muhammad.shafique\}@tuwien.ac.at}
    \IEEEauthorblockA{\IEEEauthorrefmark{2}Faculty of Information Technology, IT4Innovations Centre of Excellence, Brno University of Technology, Czech Republic
    \\\{mrazek, vasicek, sekanina\}@fit.vutbr.cz}
}

% \and
% \and
% \and
% \and
% \and

% \pagestyle{fancy}
% \lhead{Accepted for publication at the Design Automation Conference 2020 (DAC'20), San Francisco, California, USA}

\fancypagestyle{firstpage}{% Page style for first page
  \fancyhf{}% Clear header/footer
  \fancyhead[C]{To appear at the 57th Design Automation Conference (DAC), July 2020, San Francisco, CA, USA.}% Header
  \fancyfoot[C]{\thepage}% Footer
}

\maketitle

\thispagestyle{firstpage}
\pagestyle{plain}

\begin{abstract}
%%% Lukas: Always use Pareto instead of pareto!
There has been abundant research on the development of Approximate Circuits (ACs) for ASICs. 
However, previous studies have illustrated that ASIC-based ACs offer asymmetrical gains in FPGA-based accelerators. 
Therefore, an AC that might be pareto-optimal for ASICs might not be pareto-optimal for FPGAs. 
In this work, we present the \textit{ApproxFPGAs} methodology that uses machine learning models to reduce the exploration time for analyzing the state-of-the-art ASIC-based ACs to determine the set of pareto-optimal FPGA-based ACs. 
We also perform a case-study to illustrate the benefits obtained by deploying these pareto-optimal FPGA-based ACs in a state-of-the-art automation framework to systematically generate pareto-optimal approximate accelerators that can be deployed in FPGA-based systems to achieve high performance or low-power consumption. 
%\textcolor{red}{The RTL and software models of FPGA-ACs are available as open-source at \textcolor{blue}{\url{https://github.com/ehw-fit/approx-fpgas}}.}
\end{abstract}

\bstctlcite{IEEEexample:BSTcontrol}

\section{Introduction}
\label{sec:Introduction}

\noindent Field Programmable Gate Arrays (FPGAs) have become increasingly popular since their introduction in $1984$~\cite{trimberger2018three}.
Due to their (partial) run-time reconfigurability, short time-to-market, and lower prototype costs, as compared to Application-Specific Integrated Circuits (ASICs), FPGAs are preferred in a wide variety of applications.
These comprise domains like high-performance computing clusters and server platforms that offer ``\textit{FPGAs as a Service}'', and embedded and cyber-physical systems, which perform complex data-computations on the configurable arrays~\cite{watanabe2019implementation}.
The current generation of FPGAs are equipped with a wide range of capabilities that can be used to design a \textit{Programmable System} (on a Chip) by including hard IPs (IC realization) of the low-power ARM A9 processor core and other commonly used hardware accelerators, such as video codecs~\cite{crockett2014zynq}.
However, FPGAs are low-performance, power-hungry devices that are a lot less energy-efficient when compared to ASICs.

The \textit{Approximate Computing} paradigm offers a direction of research, in which the intermediate computational units can be approximated without ``significantly'' degrading the output quality, to obtain savings in power/energy consumption and latency~\cite{chippa2013analysis}.
This \textit{quality} of error-tolerance is exhibited by applications in the fields of recognition, mining, and synthesis, due to the following four factors:
\begin{inlinelist}
    \item redundancy in the processed data, 
    \item algorithms with error attenuating patterns,
    \item non-existence of a unique golden output, and 
    \item imperceptible differences in the output quality by end-users.
\end{inlinelist}
Since its re-emergence, plenty of research works from academia and industry have exploited this phenomenon across the hardware~\cite{jiang2015comparative,jiang2016comparative,mittal2016survey,hashemi2015drum,saadat2019approximate,saadat2018minimally,venkataramani2013quality,sampson2011enerj,prabakaran2018demas,echavarria2016fau,ullah2018smapproxlib,ullah2018area} and software~\cite{mishra2014iact,baek2010green,khudia2015rumba,yazdanbakhsh2015axilog} layers to obtain power/energy/latency savings.

Most of the current works on approximate circuits (AC) primarily focus on obtaining energy/power/latency savings in ASIC-based systems.
Previous studies have illustrated that ASIC-based approximate computing principles and techniques offer asymmetric savings when implemented on FPGAs~\cite{prabakaran2018demas}~\cite{ullah2018smapproxlib}~\cite{ullah2018area}.
State-of-the-art ACs for adders and multipliers can offer up to $70$\% savings in energy when synthesized for ASICs.
Whereas these designs offer minimal/asymmetric savings or at times negative savings, \textit{i.e.}, an increase in resources when synthesized for FPGAs.
This is primarily due to the architectural differences between ASICs and FPGAs.
The required functionality is realized using logic gates in ASICs and using Lookup Tables (LUTs) made of SRAM elements in FPGAs.
Therefore, an AC that offers significant savings and introduces the least error (pareto-optimal) for ASICs, might not necessarily be pareto-optimal for FPGAs.
Note, by pareto-optimal approximate circuits we mean the set of all circuits that are not dominated by any other circuit from the set of circuits in the library in terms of the evaluation metrics.

Furthermore, the works presented in ~\cite{echavarria2016fau,prabakaran2018demas,ullah2018area,ullah2018smapproxlib} have developed FPGA-based approximate circuits by analyzing the architecture of the target FPGA.
\textit{These techniques are typically not scalable, due to their manual lookup table optimizations and approximations, and do not offer multiple pareto-optimal design points} that trade-off between power consumption and introduced error.
To further illustrate these behavioral differences between ASICs and FPGAs, we present a motivational analysis of our work in the next sub-section.

\subsection{Motivational Analysis}

\begin{figure}[b]
    \centering
    \vspace{-1em}
    \includegraphics[width=\columnwidth]{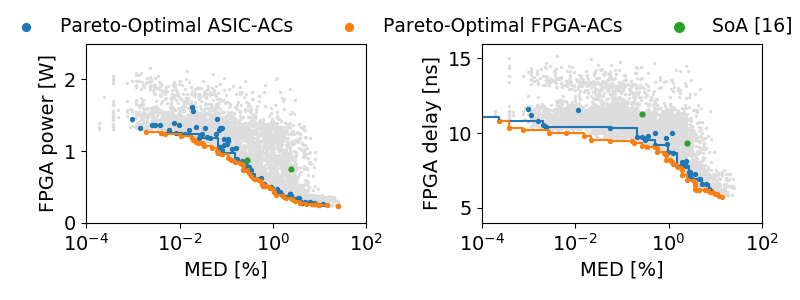}
    \caption{Analysis of Pareto-optimal Approximate Circuits for Approximate $8\text{x}8$ Multipliers and State-of-the-Art (SoA) FPGA-Based Approximate Multipliers~\cite{ullah2018area}.}
    \label{fig:asic_fpga_pareto}
\end{figure}

\noindent We synthesize and implement a small subset of $4,494$ $8\text{x}8$ unsigned approximate multiplier designs from the library of evolutionary approximate arithmetic circuits~\cite{mrazek2017evoapprox8b} and the state-of-the-art FPGA-based approximate multiplier designs~\cite{ullah2018area}.
These circuits were synthesized and implemented for the Xilinx \texttt{xc7vx485tffg1157-1} FPGA using the Vivado $2017.2$ tool-chain, with zero Digital Signal Processing blocks enabled to ensure that the designs are mapped to the reconfigurable logic (see details in Section~\ref{sec:ES}).
We also evaluate the output quality of these approximate circuits with the help of their behavioral models by computing their \textit{Mean Error Distance} (MED), which we define as the average of the absolute error difference across all the input combinations relative to the maximum number of outputs~\cite{han2013approximate}.
Based on the resources required for each of these designs and their MED, we extract the pareto-front of approximate $8\text{x}8$ multipliers for the target FPGA and compare them to the pareto-front obtained when the same designs are synthesized for ASICs.
The results of these experiments are presented in Fig.~\ref{fig:asic_fpga_pareto}.
From these results, we make the following \textbf{\textit{key observations}}:
\begin{enumerate}[label=(\arabic*),leftmargin=*]
    \item The ACs that are pareto-optimal for ASICs (ASIC-ACs) are not necessarily pareto-optimal for FPGAs (FPGA-ACs).
    As discussed earlier, this is primarily due to the differences in realizing the logic functions across the ASIC and FPGA platforms.
    \item The time required for synthesizing only $10$\% of the approximate $8\text{x}8$ multiplier library is \texttildelow$6$ days.
    This huge time requirement can also be attributed to the architectural differences between FPGAs and ASICs.
    The synthesis and routing algorithms of an FPGA tool-flow need to map the functionality to existing hardware blocks on the target FPGA while optimizing for various factors and constraints to maximize performance.
    \item State-of-the-art FPGA-based approximate multipliers~\cite{ullah2018area} are not pareto-optimal when compared to the $10$\% subset of approximate multipliers from~\cite{mrazek2017evoapprox8b}.
    Similarly, the other FPGA-based approximate adders and multipliers presented in~\cite{prabakaran2018demas}~\cite{ullah2018smapproxlib} are neither pareto-optimal nor scalable.
    Due to their manual optimizations and circuit designs, they are not effective in achieving similar performance/power trade-offs, as illustrated by the evolutionary approximate arithmetic library for larger bit-widths.
\end{enumerate}

% \begin{itemize}
%     \item Show that synthesis results for ASIC-based 8x8 approximate multipliers from the EvoApprox library on the FPGA platform.
%     \item analyze the benefits obtained by the system and compare them with the benefits of ASICs
%     \item show for the area, power, latency and PDP (Power Delay Product)
%     \item clearly shows the set of pareto-optimal points for FPGAs is different from the pareto-optimal points for ASICs.
%     \item Present observations of the results and then present the research challenges.
% \end{itemize}

Based on these observations, we have identified the following \textbf{\textit{research challenges}}:
% \subsection{Research Challenges}
\begin{itemize} [leftmargin=*]
    \item Based on the time required for synthesizing and implementing a small subset of the designs for the target FPGA, the time required for exhaustively exploring all designs in the data-set would be in the magnitude of $100$s of hours, or a couple of weeks.
    \begin{itemize}
        \item \textit{How to efficiently reduce the time required for exploring the design-space of approximate arithmetic units in FPGAs?}
        \item \textit{Can we explore the concepts of machine learning in order to reduce the exploration time by estimating FPGA parameters? If yes, which machine learning algorithm?}
    \end{itemize}
    \item There is a nonexistence of pareto-optimal FPGA-ACs, which can offer a design-space trade-off between the resources consumed and the error introduced.
    \begin{itemize}
        \item \textit{How can we determine a set of pareto-optimal FPGA-ACs that can be deployed in error-tolerant applications to obtain power/energy/latency savings?}
    \end{itemize}
    \item Unavailability of a systematic automation framework that can be used to develop FPGA-ACs for a given error-tolerant application and its quality requirements.
    \begin{itemize}
        \item \textit{How to systematically deploy the FPGA-ACs in a given error tolerant application to maximize performance or power/energy savings?}
    \end{itemize}
\end{itemize}

To address these research challenges, we propose the following \textbf{\textit{novel contributions}}:
% \subsection{Novel Contributions}
\begin{itemize}[leftmargin=*]
    \item We propose the \textit{ApproxFPGAs} methodology that deploys machine learning (ML) models, which can be used to estimate the power and latency of the approximate circuits.
    These ML models are trained using a  small subset of the evolutionary approximate circuits~\cite{mrazek2017evoapprox8b}.
    \item Based on the estimates, we propose to construct a pseudo-pareto-front, which can be used to determine the set of pseudo-pareto-optimal approximate circuits for varying bit-widths of the approximate arithmetic units.
    These models can then be subsequently synthesized for the target FPGA to measure the exact power and latency of these FPGA-ACs.
    \item These pareto-optimal FPGA-ACs are open-source and available online at \textcolor{blue}{\url{https://github.com/ehw-fit/approx-fpgas}}, to enable reproducible research and foster development in this area.
    \item We also perform a case-study by deploying these pareto-optimal FPGA-ACs in a state-of-the-art automation framework that can systematically generate approximate accelerators, which can be deployed in FPGA-based systems to achieve high-performance and/or low power/energy consumption.
\end{itemize}

\section{The ApproxFPGAs Methodology}
\label{sec:ApproxFPGAs}

\begin{figure}[h!]
    \centering
    \includegraphics[width=\columnwidth]{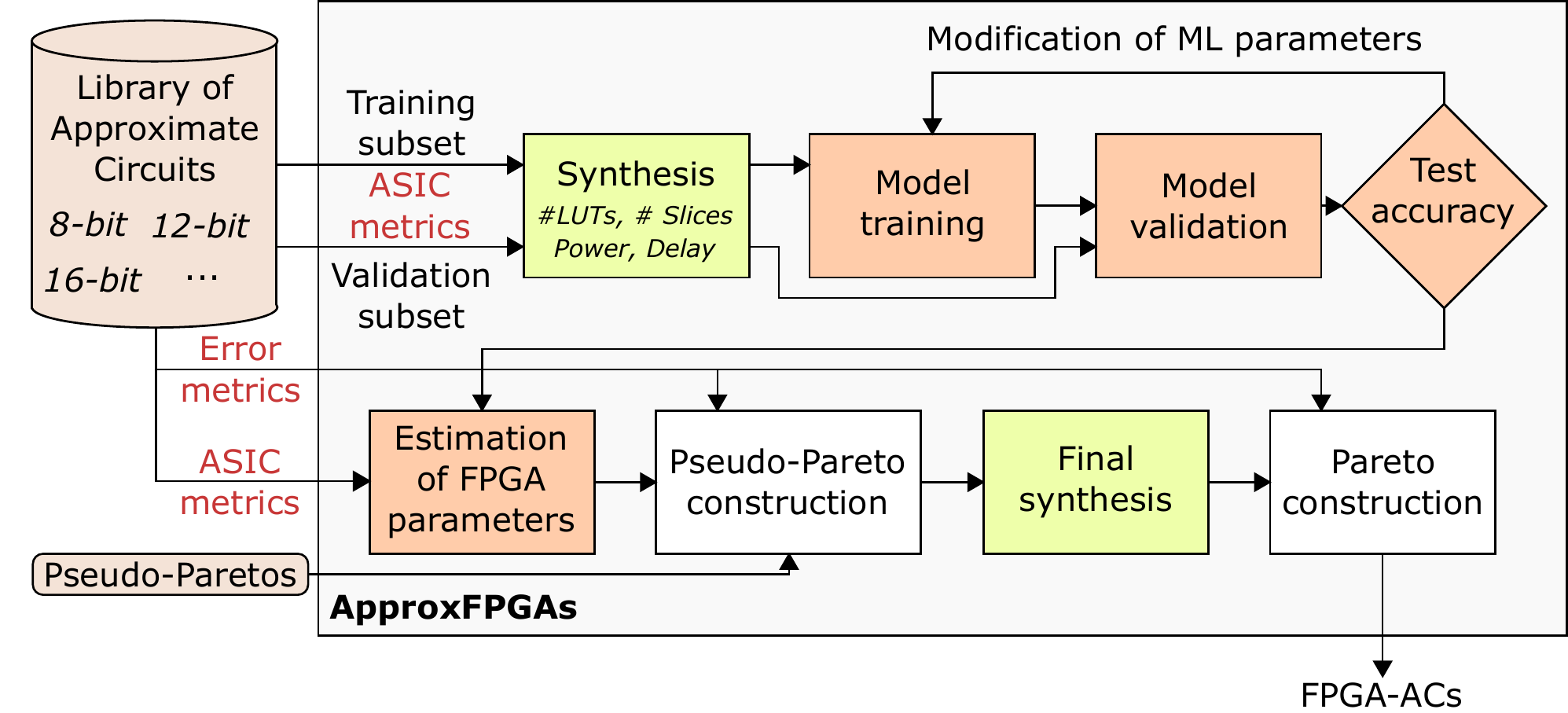}
    \caption{An Overview of the ApproxFPGAs Methodology}
    \label{fig:ApproxFPGAs}
\end{figure}

\noindent \textbf{Overview:} Fig.~\ref{fig:ApproxFPGAs} presents an overview of the proposed methodology.
The complete procedure can be divided into two sub-parts,
\begin{inlinelist}
    \item the first part deals with the training and testing of the ML models, which can be used to efficiently estimate the hardware resources of a given approximate arithmetic design, while
    \item the second part deals with the construction of the pareto-optimal FPGA-ACs, which can be deployed in error-tolerant applications.
\end{inlinelist}

\textbf{Inputs:} We start by compiling the library of approximate arithmetic circuits that need to be analyzed and deployed in the target application.
Without loss of generality, in this work, we consider the evolutionary library of approximate adder and multiplier circuits for illustrating the benefits of our methodology~\cite{mrazek2017evoapprox8b}.
Note, the use of other state-of-the-art designs is orthogonal to our approach and they can be appropriately included, with necessary modifications, in the library of approximate circuits.

\begin{figure}[b]
    \centering
    \includegraphics[width=\columnwidth]{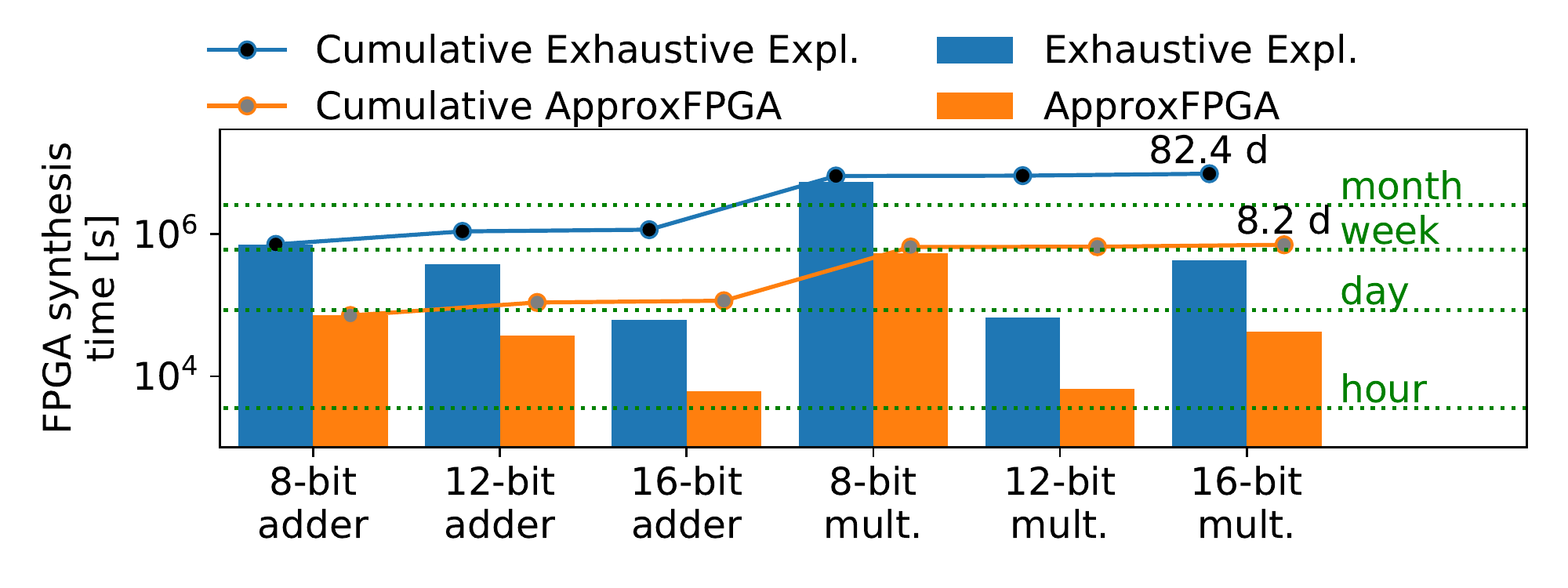}
    \caption{Time Required for Exhaustive Exploration Compared to our \textit{ApproxFPGAs} Approach for all ACs in the Library.}
    \label{fig:ExhaustiveExploration}
\end{figure}

\textbf{Exhaustive Exploration:} Due to the large number of designs present in the library, the time required for exploring all the designs, exhaustively, might be quite large, as initially stated in Section~\ref{sec:Introduction}.
Fig.~\ref{fig:ExhaustiveExploration} presents a brief illustration of the estimated time required for synthesizing all the approximate circuits present in the library for the target FPGA.
As can be observed, when the number of ACs in the library increases, the time required for exploring the designs rises and reaches a magnitude of ~$100$s of hours.
Therefore, exhaustive exploration is not a feasible option for identifying the pareto-optimal approximate circuits for FPGAs.
Fig.~\ref{fig:ExhaustiveExploration} also illustrates the savings in exploration time when the proposed \textit{ApproxFPGAs} methodology is used for exploration as opposed to exhaustive exploration. 
The exploration time is reduced by a factor of \texttildelow$10\times$ from $82.4$ days to $8.2$ days, including the time required for synthesizing the data-set, training and evaluating the ML models, and re-synthesizing the pareto-optimal FPGA-ACs.

\textbf{ML-Model Learning:} Due to the infeasible time requirements of exhaustive exploration, we propose to train and evaluate a wide variety of statistical and machine learning (S/ML) models, which can be used to estimate the resource requirements of an approximate circuit, given its hardware description.
These S/ML models can be used to estimate FPGA parameters like power consumption ($W$), latency ($ns$), and area (\#$LUTs$).
Training these models requires a labeled data-set, with the FPGA parameters as the output labels and the hardware description of the AC as the input data.
We build this data-set by randomly extracting a $10$\% subset of the complete library of ACs and synthesizing them for the target FPGA platform. 
This subset is further partitioned into training ($80$\%) and validation ($20$\%) data-sets, which are then used to train and evaluate the various machine learning models, respectively.
Without loss of generality, in this work, we evaluate the applicability of the most-commonly used light-weight S/ML models (see Table~\ref{tab:SML}) to reduce the time required for exploring the library of ACs.
We iteratively evaluate the accuracy of the models and modify their parameters based on the correlation obtained on the validation data-set to further improve the model's accuracy.
Instead of synthesizing and implementing each circuit in the library, which might take weeks to months, we can roughly estimate the FPGA parameters of all circuits using these models in the order of seconds.
To estimate the accuracy of these ML models, we propose the \textit{fidelity} metric, which evaluates the relationship between the measured (\texttt{mes}) and estimated (\texttt{est}) FPGA parameters for any two ACs in the library.
We compute the \textit{fidelity} ($F$) of a set of ACs, $X$, as: %between two ACs $x$ and $y$ as:

\begin{equation}
    F(X) = \frac{\sum_{x_1 \in X}\sum_{x_2 \in X}E(x_1, x_2)}{|X|^{2}}
\end{equation}

where $E$ denotes the correctness of the relationship between the estimated and measured FPGA parameters:
\begin{equation}
    E(x,y) = \Bigg\{
    \begin{matrix}
        1 & \text{If \texttt{est}($x$) $\mathcal{R}$ \texttt{est}($y$)} \\
          & \wedge \text{ \texttt{mes}($x$) $\mathcal{R}$ \texttt{mes}($y$)} \\
        0 & \text{Otherwise}
    \end{matrix}
\end{equation}
where $\mathcal{R}$ denotes one of the following relations $\{<,>,=\}$ between the FPGA parameters of the ACs.
Due to their availability

\begin{table}[t]
\centering
\caption{\protect\centering\textbf{List of Light-weight Statistical/Machine Learning Models Used in ApproxFPGAs.}}
\label{tab:SML}
\begin{tabular}{c|L{2.2cm}|}
\cline{2-2}
                           & \multicolumn{1}{c|}{\textbf{Statistical/ML Model}}                \\ \hline
\multicolumn{1}{|c|}{ML1}  & Regression \textit{w.r.t} ASIC-AC Power   \\ \hline
\multicolumn{1}{|c|}{ML2}  & Regression \textit{w.r.t} ASIC-AC Latency \\ \hline
\multicolumn{1}{|c|}{ML3}  & Regression \textit{w.r.t} ASIC-AC Area    \\ \hline
\multicolumn{1}{|c|}{ML4}  & PLS Regression                                             \\ \hline
\multicolumn{1}{|c|}{ML5}  & Random Forest                                              \\ \hline
\multicolumn{1}{|c|}{ML6}  & Gradient Boosting                                          \\ \hline
\multicolumn{1}{|c|}{ML7}  & Adaptive Boosting (AdaBoost)                               \\ \hline
\multicolumn{1}{|c|}{ML8}  & Gaussian Process                                           \\ \hline
\multicolumn{1}{|c|}{ML9}  & Symbolic Regression                                        \\ \hline
% \multicolumn{1}{|c|}{ML10} & Kernel Ridge                                               \\ \hline
% \multicolumn{1}{|c|}{ML11} & Bayesian Ridge                                             \\ \hline
% \multicolumn{1}{|c|}{ML12} & Coordinate Descent (Lasso)                                 \\ \hline
% \multicolumn{1}{|c|}{ML13} & Least Angle Regression (LARS)                              \\ \hline
% \multicolumn{1}{|c|}{ML14} & Ridge Regression                                           \\ \hline
% \multicolumn{1}{|c|}{ML15} & Stochastic Gradient Descent                                \\ \hline
% \multicolumn{1}{|c|}{ML16} & K-Nearest Neighbours                                       \\ \hline
% \multicolumn{1}{|c|}{ML17} & Multi-Layer Perceptron (MLP)                               \\ \hline
% \multicolumn{1}{|c|}{ML18} & Decision Tree                                              \\ \hline
\end{tabular}
\quad
\begin{tabular}{c|L{2.2cm}|}
\cline{2-2}
                           & \multicolumn{1}{c|}{\textbf{Statistical/ML Model}}                \\ \hline
% \multicolumn{1}{|c|}{ML1}  & Regression \textit{w.r.t} ASIC-AC Power   \\ \hline
% \multicolumn{1}{|c|}{ML2}  & Regression \textit{w.r.t} ASIC-AC Latency \\ \hline
% \multicolumn{1}{|c|}{ML3}  & Regression \textit{w.r.t} ASIC-AC Area    \\ \hline
% \multicolumn{1}{|c|}{ML4}  & PLS Regression                                             \\ \hline
% \multicolumn{1}{|c|}{ML5}  & Random Forest                                              \\ \hline
% \multicolumn{1}{|c|}{ML6}  & Gradient Boosting                                          \\ \hline
% \multicolumn{1}{|c|}{ML7}  & Adaptive Boosting (AdaBoost)                               \\ \hline
% \multicolumn{1}{|c|}{ML8}  & Gaussian Process                                           \\ \hline
% \multicolumn{1}{|c|}{ML9}  & Symbolic Regression                                        \\ \hline
\multicolumn{1}{|c|}{ML10} & Kernel Ridge                                               \\ \hline
\multicolumn{1}{|c|}{ML11} & Bayesian Ridge                                             \\ \hline
\multicolumn{1}{|c|}{ML12} & Coordinate Descent (Lasso)                                 \\ \hline
\multicolumn{1}{|c|}{ML13} & Least Angle Regression                                     \\ \hline
\multicolumn{1}{|c|}{ML14} & Ridge Regression                                           \\ \hline
\multicolumn{1}{|c|}{ML15} & Stochastic Gradient Descent                                \\ \hline
\multicolumn{1}{|c|}{ML16} & K-Nearest Neighbours                                       \\ \hline
\multicolumn{1}{|c|}{ML17} & Multi-Layer Perceptron (MLP)                               \\ \hline
\multicolumn{1}{|c|}{ML18} & Decision Tree                                              \\ \hline
\end{tabular}
\end{table}

\textbf{Pareto Construction:} Based on the outcome of our experiments (see Section~\ref{sec:Results}), we select the best S/ML models to estimate the FPGA parameters of all ACs in the library.
Based on these parameter-estimates, we can determine the pareto-optimal FPGA-ACs.
% For such parameters, new pareto fronts can be identified in the library. 
However, we have observed that these models have limited fidelity, because of which the real pareto-optimal ACs can be dominated by the ACs where the estimation was incorrect.
Therefore we propose to construct multiple pseudo-pareto-fronts from the input set (library) of ACs $C$. 
We determine the first set of pseudo-pareto-optimal ACs ($F_1$) from the initial set of all ACs $C$. 
Next, we eliminate all these pseudo-pareto-points from the input set to construct the second pseudo-pareto-front, i.e., using $C \setminus F_1$ as the input, we determine $F_2$.
Similarly, we construct the third pseudo-pareto-front $F_3$, using the input $C \setminus (F_1 \cup F_2)$, and so on.
By constructing multiple pseudo-pareto-fronts, we mitigate the inaccuracies associated with our S/ML models.
The ACs lying on these pseudo-pareto-fronts can be subsequently synthesized again using our work-flow to determine the accurate FPGA parameters and the resources required.
Hence, we have to synthesize an additional number of ACs when we are constructing multiple pseudo-pareto-fronts.

Based on the real FPGA parameter measurements obtained from the synthesis and implementation reports of Vivado, we construct an open-source library of pareto-optimal FPGA-ACs that offers a trade-off between the output quality and the resources consumed. 
This library can be subsequently utilized by application and system developers, to further maximize performance or power and energy savings obtained while satisfying the quality constraints of the application.
The RTL and behavioral models of the FPGA-ACs are open-source and are available online at \textcolor{blue}{\url{https://github.com/ehw-fit/approx-fpgas}}.

%\todored{TODO: The technical details of the pseudo-pareto construction, final synthesis, and the actual pareto construction need to be included here. Could you please include the same?}

%Based on the parameter estimates obtained from the S/ML models, we construct a pseudo-pareto-front of FPGA-ACs.

% \subsection{FPGA-AutoAx}

\textbf{AutoAx-FPGA:} To incorporate the set of pareto-optimal FPGA-ACs in different error-tolerant applications, we modify the state-of-the-art AutoAx~\cite{mrazek2019autoax} framework to include the functionality of designing ACs for a given application that can be deployed in FPGA-based systems.
The traditional AutoAx framework searches the design-space of approximate components to select and combine approximation components, in order to generate an approximate hardware accelerator that maximizes the energy savings.
Initially, a set of random approximation assignments are evaluated for the target accelerator circuit, to get the quality of results (QoR) and hardware (HW) cost of the accelerator.
Based on these values, QoR and HW cost estimators are constructed, which can be used to explore the complete design-space of approximate components for the given accelerator and to determine the set of pareto-optimal circuits for the given application.
To generate approximate accelerators for a given application, which can be used in low-power and/or high-performance FPGA-based systems, we propose to include the following functionality in AutoAx:
\begin{inlinelist}
    \item we replace the library of pareto-optimal ASIC-ACs with the set of pareto-optimal FPGA-ACs obtained from the proposed \textit{ApproxFPGAs} methodology,
    \item we modify the estimators used in AutoAx to estimate the FPGA parameters of the approximate accelerator instead of their ASIC-based HW costs.
\end{inlinelist}

% (i), and (ii)  The research questions are (i) is it better to use FPGA-based optimization than ASIC-based if we optimize for FPGAs? and (ii) can we successfully estimate all FPGA parameters of the target accelerator?

% The HW cost was estimated of area, delay and power consumption of the fully characterized single components from the library assigned to the operations. 

% \begin{itemize}
%     \item Figure that shows an overview of the methodology
%     \item Brief Explanation of each step in the methodology
%     \item Input to the methodology: library of ASIC-based approximate arithmetic circuits, FPGA hardware, and configuration
%     \item Synthesis and implementation of exhaustively synthesizing all designs in the dataset is highly time and resource consuming.
%     \item Use of machine learning models to reduce the exploration time by synthesizing and implementing a small portion of the data-set, which can be used to train and test the ML models for the target FPGA platform.
%     \item Based on the models obtained we explore the dataset using genetic algorithms such as NSGA-II to offer a 60-70\% coverage of pareto-points when compared to exhaustive search.
%     \item Modifying the AutoAx framework to include the functionality of systematically developing approximate circuits for FPGAs
%     \item \todored{TODO: detail all aspects of the framework}
% \end{itemize}
\section{Experimental Setup}
\label{sec:ES}

\setcounter{figure}{5}
\begin{figure*}[b]
    \centering
    \includegraphics[width=\textwidth]{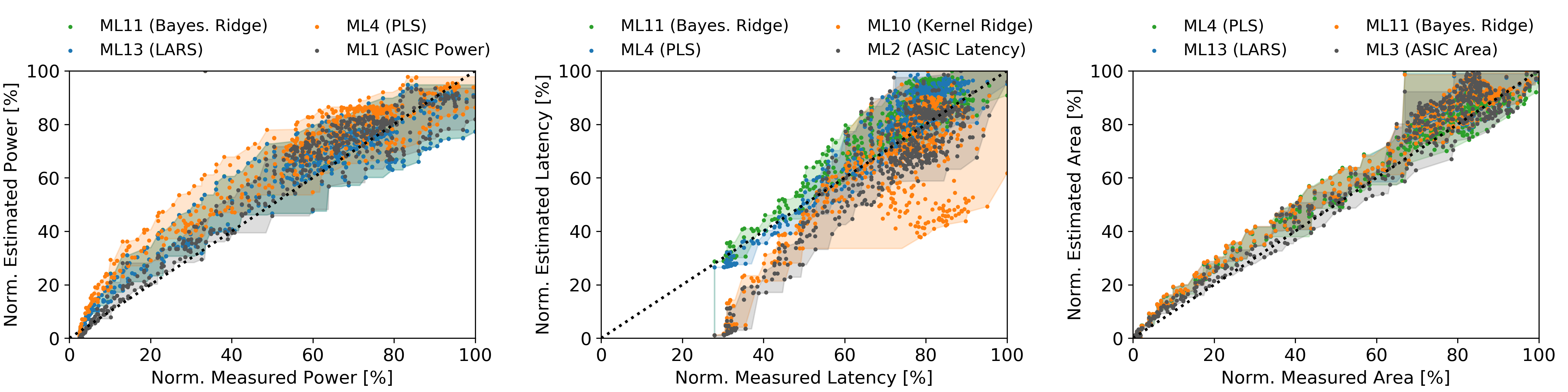}
    \caption{Correlation Analysis of the the Top-$3$ S/ML Techniques for the Library of $16\text{x}16$ Approximate Multipliers.}
    \label{fig:est:corr}
\end{figure*}

\setcounter{figure}{4}
\begin{figure}[b]
    \centering
    % \vspace{-2em}
    \includegraphics[width=\columnwidth]{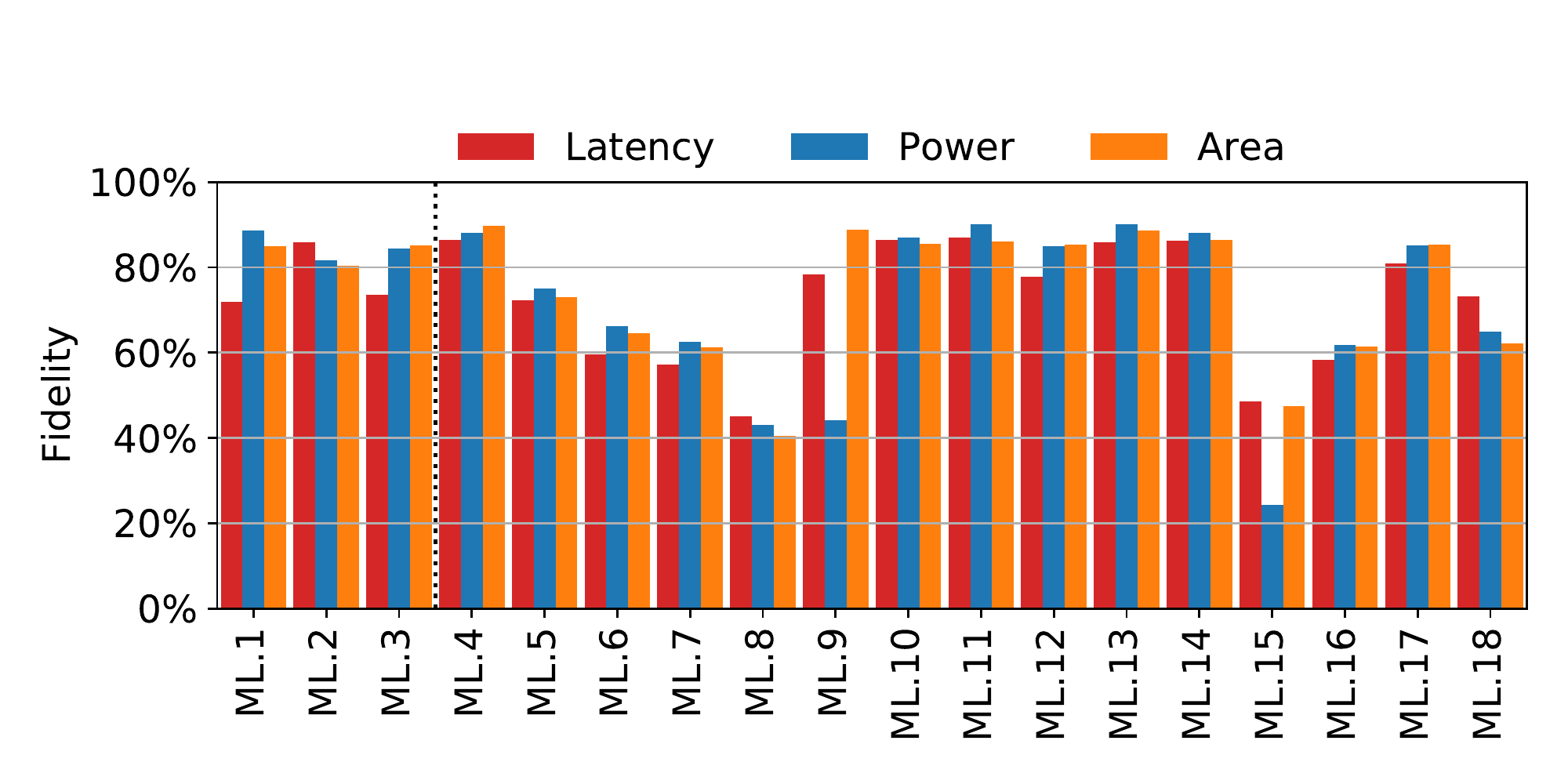}
    \caption{Fidelity Analysis of the $3$ FPGA parameters for the Different S/ML Techniques Described in Table~\ref{tab:SML}.}
    \label{fig:fidelity}
\end{figure}

\noindent The RTL (in Verilog) and behavioral models (in C) of the evolutionary approximate arithmetic circuits are open-source and readily accessible\footnote{https://github.com/ehw-fit/evoapproxlib}.
These designs are synthesized and implemented (i.e., place \& route) using the Vivado Design Suite $2017.2$ for the target FPGA \texttt{xc7vx485tffg1157-1}, to extract their area, power, and timing reports.
We restrict the placement and routing algorithms of the Xilinx Vivado by disabling the use of the FPGA's DSP logic blocks. We do this to ensure that the designs are mapped on to the configurable logic.
These reports are used to extract the FPGA parameters, which are subsequently used for training and evaluating the S/ML models.
The S/ML models are implemented, trained, and tested inside the Python $3.7$ environment with the help of the \textit{scikit-learn} library.
The RTL designs were synthesized on an Intel Core i$5-7600$ CPU with $16$GB of internal memory and a $256$GB Solid-State Drive (SSD).
The S/ML models were trained and evaluated on an Intel Xeon CPU E$5-2630$ with $16$GB of internal memory.
An overview of our work-flow is presented in Fig.~\ref{fig:ES}.

\setcounter{figure}{3}
\begin{figure}[H]
    \centering
    \includegraphics[width=\columnwidth]{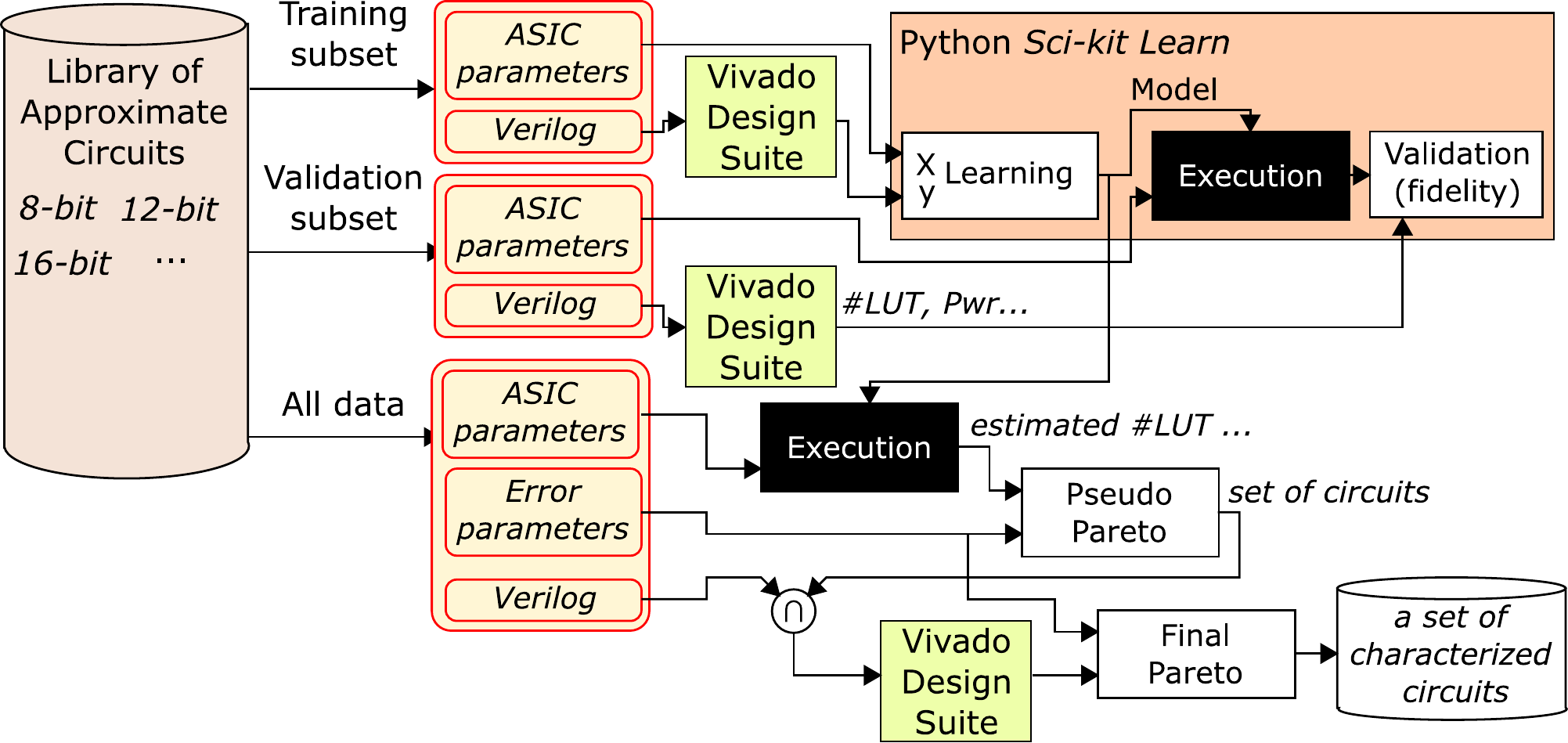}
    \caption{Overview of Our Experimental Work-flow.}
    \label{fig:ES}
\end{figure}

% \begin{itemize}
%     \item Python
%     \item ML environment
%     \item Xilinx Vivado 2019
%     \item Xilinx Zynq 7000 series
%     \item Target Platform:
%     \item Hardware platform for synthesis
%     \item hardware platform for testing the ML models
%     \item Overview figure
% \end{itemize}
\section{Results \& Discussion}
\label{sec:Results}

\noindent\textbf{Fidelity:}
First, we illustrate the accuracy of the $18$ S/ML models that we have evaluated inside our \textit{ApproxFPGAs} framework. 
We do this by studying the fidelity of these models with respect to the three important FPGA parameters, namely, latency ($ns$), power ($mW$), and area (\#$LUTs$).
The fidelity of these models is evaluated on the validation data-set.
The results of these experiments are presented in Fig.~\ref{fig:fidelity}.
\textit{From these results, we make the following} \textbf{\textit{key observations}}:
\begin{itemize}[leftmargin=*]
    \item Tree-based methods, like \textit{Decision Trees} and \textit{Random Forrest}, achieve above-average accuracy in estimating the FPGA parameters and retaining their relationship to the other ACs.
    \item Based on further analysis, we also observed that generalization of models across all bit-widths is not very effective, i.e., estimating FPGA parameters of higher bit-width ($12$-/$16$-bit) designs (adder or multiplier) using a model learned from a lower bit-width ($8$-bit) designs is not very effective. 
    On average, we observed that the fidelity of the higher bit-width designs decreased from $88$\% to $53$\% when using models trained with lower bit-width designs as opposed to designs of the same bit-width.
    \item Ridge models such as \textit{Kernel Ridge} and \textit{Bayesian Ridge}, typically, illustrate the best fidelity.
\end{itemize}

\setcounter{figure}{6}
\begin{figure*}[t]
    \centering
    \includegraphics[width=\textwidth]{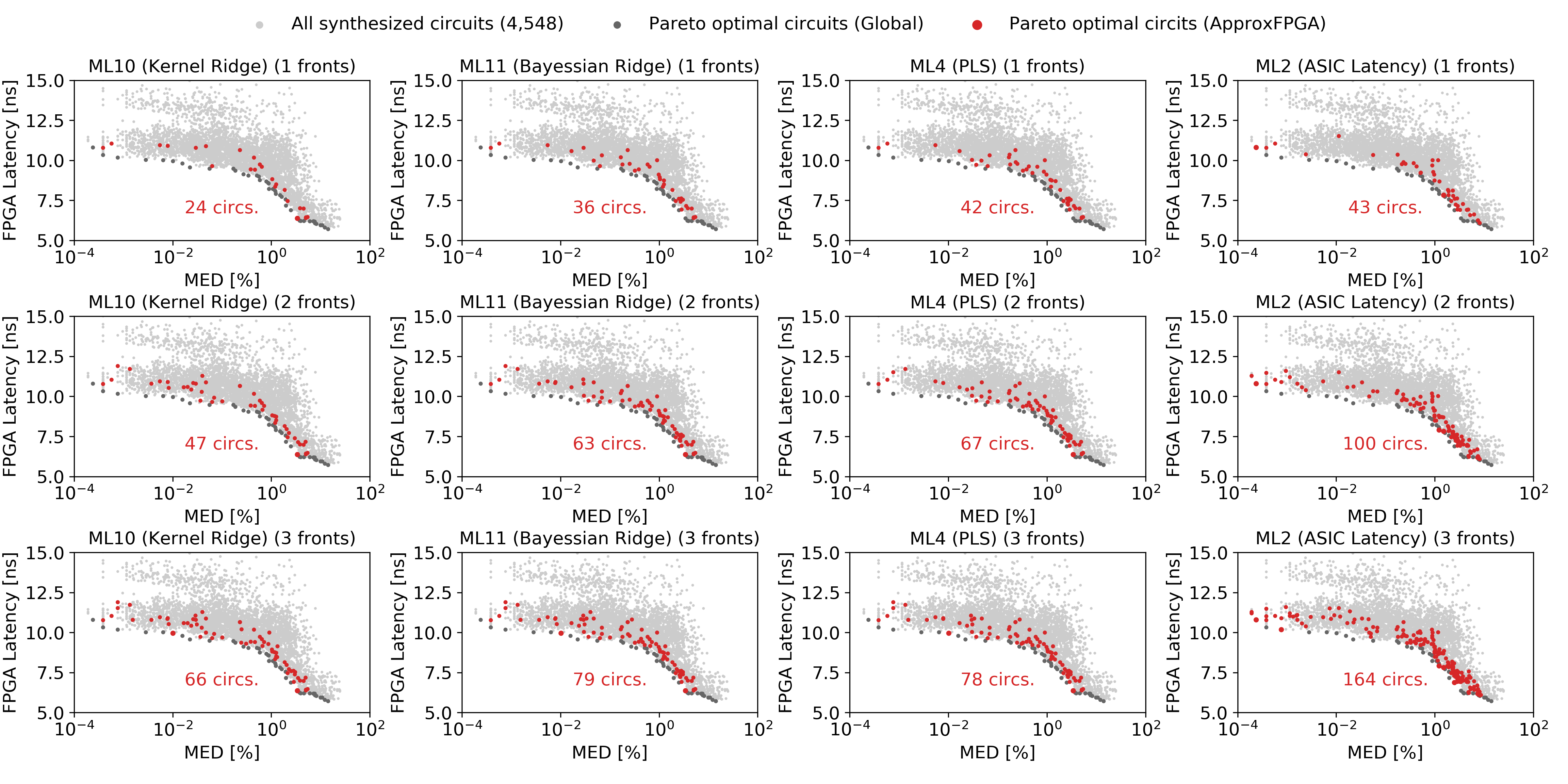} % there are to many points in PDF, I suggest to use PNG instead
    \caption{Analysis of Constructing Multiple Pareto-fronts for the $8\text{x}8$ Approximate Multiplier Library \textit{w.r.t} FPGA Latency.}
    \label{fig:pareto_est}
\end{figure*}

\begin{figure*}[b]
    \centering
    \includegraphics[width=\textwidth]{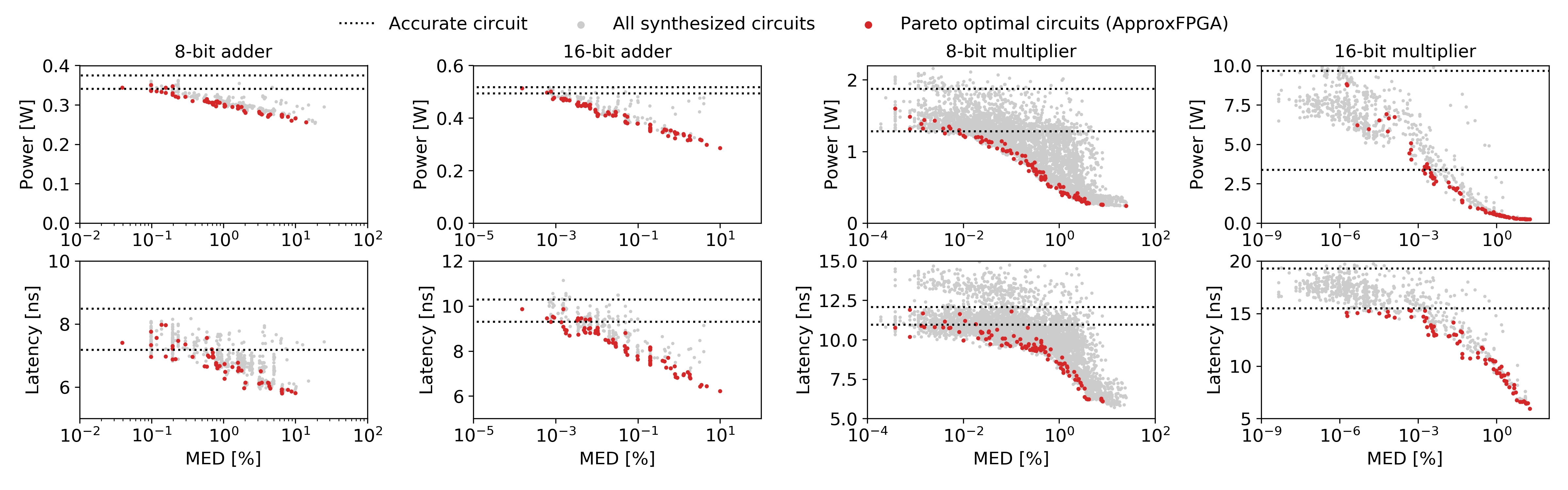}
    \caption{Evaluation of the Pareto-optimal FPGA-ACs Obtained using the \textit{ApproxFPGAs} Methodology.}
    \label{fig:circuits}
\end{figure*}

We also summarize the top-$3$ S/ML models for each FPGA parameter, along with the fidelity achieved for each case, in Table~\ref{tab:fidelity}. Likewise, we identify the models that achieve maximum fidelity when obtained using regression analysis on their corresponding ASIC parameters.

\begin{table}[ht]
    \centering
    \caption{Fidelity of the top-$3$ ML Models for the Estimating the FPGA parameters} 
    \label{tab:fidelity}
    \begin{tabular}{| c c | c c | c c |}\hline
\multicolumn{2}{|c|}{\bf FPGA Latency} &    \multicolumn{2}{c|}{\bf FPGA Power} & \multicolumn{2}{c|}{\bf FPGA Area} \\\hline
Model & Fidelity & Model & Fidelity & Model & Fidelity \\\hline
ML11 &    $90$\%    & ML11 & $91$\% & ML4 & $89$\% \\
ML4 & $89$\% & ML13 & $91$\% & ML13 & $88$\% \\
ML10 & $87$\% & ML4 & $89$\% & ML11 & $86$\% \\\hline
ML2 & $89$\% & ML1 & $90$\% & ML3 & $84$\% \\\hline
    \end{tabular}
\end{table}

\textbf{Correlation of ML Models:}
Next, we illustrate the correlation between the estimated FPGA parameters and their measured values when the top-$3$ S/ML models are used on the library of approximate $16\text{x}16$ multipliers.
These results are illustrated in Fig.~\ref{fig:est:corr}.
From these results, we make the following \textbf{\textit{key observations}}:
\begin{itemize}[leftmargin=*]
    \item The \textit{Bayesian Ridge} and \textit{PLS regression} techniques can be used as standalone techniques to estimate all three FPGA parameters, as they are one of the top-$3$ models for all three parameters.
    \item Statistical regression with respect to the corresponding ASIC parameters is equally useful in estimating the FPGA parameters of the given circuit.
    \item Due to the \texttildelow$30$\% bias illustrated by the model, latency is not estimated accurately, especially using regression with ASIC parameters and \textit{Kernel Ridge}.
    This leads to a scenario where the circuit latency is under-estimated by the model, including certain pareto-optimal designs.
\end{itemize}

\textbf{Construction of the Pareto-fronts:}
As discussed earlier, we construct multiple pareto-fronts to ensure that non-pareto-optimal designs are not missed by our methodology.
Towards this, we illustrate the benefits of constructing multiple pareto-fronts sequentially for estimating the FPGA latency using the top-$3$ S/ML models and Regression with respect to ASIC latency.
Fig.~\ref{fig:pareto_est} illustrates the results of constructing $1$, $2$, and $3$ pareto-fronts using the technique discussed in Section~\ref{sec:ApproxFPGAs}.
From these results, we make the following \textbf{\textit{key observations}}:
\begin{itemize}[leftmargin=*]
    \item Using ML-based techniques for estimating the FPGA parameters reduces the total number of synthesized circuits by a factor of \texttildelow$9.9\times$ to $4,548$, including the training and validation data-set and synthesis of pseudo-pareto-optimal points, instead of synthesizing the complete library of approximate $8\text{x}8$ multipliers.
    \item The ML models are highly effective in selecting the pseudo-pareto-optimal designs that have to be re-synthesized, as compared to the regression analysis w.r.t ASIC latency, which increases the number of circuits to be explored from $79$ in \textit{Bayesian Ridge} to $164$, effectively doubling the number of new circuits to be synthesized.
    \item The best results are obtained when we effectively combine the pseudo-pareto-optimal points obtained from multiple ML models.
    Therefore, we need to consider a union of all the pareto-fronts $\bigcup_{i=1}^{n}F_i$ to determine the final set of pareto-optimal FPGA-ACs.
\end{itemize}

\textbf{Pareto-Optimal FPGA-ACs:}
Fig.~\ref{fig:circuits} illustrates the set of FPGA-ACs synthesized to obtain the subset of pareto-optimal FPGA-ACs using our proposed \textit{ApproxFPGAs} methodology on the library of $8$-, $16$-bit adders and $8\text{x}8$, $16\text{x}16$ multipliers.
Although we have not exclusively determined and synthesized all the pareto-optimal designs, we have reduced the exploration time a factor of \texttildelow$10\times$ to obtain, on average, $71$\% percentage coverage of the pareto-optimal designs present in the library of approximate circuits.
This is quite explicitly illustrated with the help of the pareto-front in the designs with a higher number of ACs present in the library, such as the approximate multipliers, and a little less explicit for libraries with a lower number of circuits, like the approximate adders.
Similarly, we have generated the pareto-optimal ACs for the $12$-bit approximate adder and $12\text{x}12$ approximate multiplier.

% Based on these results, we can also infer that the S/ML models used in the \textit{ApproxFPGAs} methodology can be trained using a small portion of the library of approximate circuits ($10$\% in our case), provided the library is exhaustive and contains a large number of designs.

\begin{figure}[ht]
    \centering
    \includegraphics[width=\columnwidth]{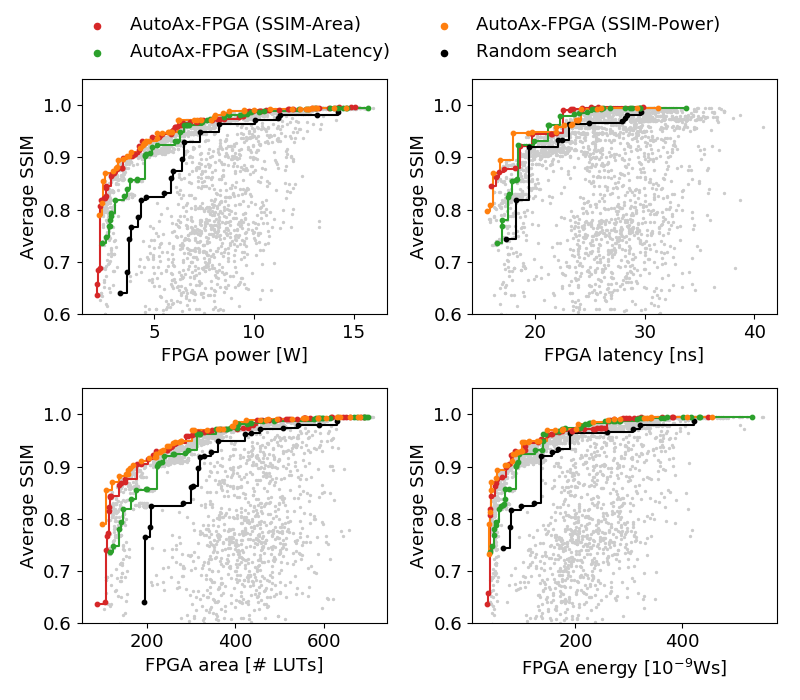}
    \caption{Analysis of the ACs from FPGA-AutoAx Compared to the Basic Random Search.}
    \label{fig:autoax:res}
\end{figure}

\textbf{AutoAx-FPGA:}
Finally, we present the results of modifying the AutoAx framework to include the functionality of generating pareto-optimal accelerators for FPGA-based systems.
We evaluated the modified AutoAx-FPGA methodology using a Gaussian Filter as a case-study and the input of $9$ pareto-optimal $8\text{x}8$ approximate multipliers and $8$ $16$-bit approximate adders.
The QoR of the Gaussian filter's output is estimated using the structural similarity index (SSIM), for which we build an estimator.
First, we generate a training and validation data-set of $5,000$ random approximate circuits for the given Gaussian filter, which was synthesized and implemented using the Vivado work-flow to measure their FPGA parameters such as area, latency, and power consumption. 
Similar to the AutoAx methodology, we constructed estimators that can determine the FPGA parameters for the other circuits in the library, and construct $3$ different pareto-fronts using the \textit{hill-climber} algorithm.
We thereby reduce the number of accelerator circuits to be explored from $4.95\times10^{14}$ to $368$, $444$, and $946$ possible designs for each of the FPGA parameter-QoR scenarios, namely, latency-SSIM, power-SSIM, and area-SSIM, respectively.
Each of these designs is synthesized in the Vivado work-flow and their behavioral models are deployed in the image processing environment to measure their FPGA parameters and determine their SSIM.
These results are illustrated in Fig.~\ref{fig:autoax:res}
We can observe that AutoAx-FPGA achieves better results when compared to a simple random search. 
Furthermore, we can also observe that the optimization for area and power improve the savings obtained in other FPGA parameters as well, which is not the case when we optimize for latency. 
For example, in the case where we optimize for latency in Fig.~\ref{fig:autoax:res}, we would expect the SSIM-Latency pareto-front to encompass the best ACs in terms of latency, but this is not the case as the latency-estimator is not very effective.
However, since the other two pareto-fronts improve the savings for other FPGA parameters as well, they outperform the SSIM-Latency pareto-front ACs, even in terms of latency.

\section{Conclusion}
\label{sec:Conclusion}

\noindent We presented the \textit{ApproxFPGAs} methodology, for embracing the use of current state-of-the-art ASIC-based approximate circuits for FPGA-based systems. 
We synthesize a partial subset of the library of arithmetic circuits to establish the training and validation data-set, which can be used to teach and evaluate the models' applicability.
Based on the outcome, we chose the top-$3$ models that achieve the best fidelity to estimate the FPGA parameters for all the circuits in the data-set, which is used to subsequently construct multiple pseudo-pareto-fronts.
The circuits on these pareto-fronts are re-synthesized to measure the correct FPGA parameters and determine the final set of pareto-optimal FPGA-ACs, which can be used by system developers and application-designers to develop low-power or high-performance FPGA-based accelerators.
This set of pareto-optimal arithmetic FPGA-ACs is open-source and available online at \textcolor{blue}{\url{https://github.com/ehw-fit/approx-fpgas}}.
Finally, we evaluate the applicability of these pareto-optimal ACs by using a modified version of the state-of-the-art AutoAx framework to illustrate the benefits obtained.

\section*{Acknowledgement} 
This work was partially supported by Doctoral College Resilient Embedded Systems which is run jointly by TU Wien's Faculty of Informatics and FH-Technikum Wien, and partially by Czech Science Foundation project 19-10137S.

\bibliographystyle{IEEEtran}
\bibliography{References}

\end{document}